\documentclass[11pt,prd,a4paper,preprintnumbers,amsmath,amssymb,nofootinbib]{article}
\pdfoutput=1
\usepackage{feynmp-auto,expdlist}
\usepackage{amsmath,amsfonts,amssymb}
\usepackage{graphicx}
\usepackage{enumerate}
\usepackage{hyperref}
\usepackage{latexsym}
\usepackage{hepnicenames}
\usepackage{enumerate}
\usepackage{soul}
\usepackage[normalem]{ulem}
\usepackage{wasysym}
\usepackage{makecell}
\usepackage{bbm}

\font\tenrsfs=rsfs10 at 12pt
\font\sevenrsfs=rsfs7
\font\fiversfs=rsfs5
\newfam\rsfsfam
\textfont\rsfsfam=\tenrsfs
\scriptfont\rsfsfam=\sevenrsfs
\scriptscriptfont\rsfsfam=\fiversfs

\numberwithin{equation}{section}

\usepackage{mathrsfs}
\usepackage{braket}
\usepackage{titling}
\usepackage{amsmath}
\usepackage{slashed}
\usepackage{amssymb}
\usepackage{epsfig}
\usepackage{graphicx}
\usepackage{color}
\usepackage{rotating}
\usepackage{hyperref}
\usepackage[margin=1.in]{geometry}
\usepackage[table,xcdraw,dvipsnames]{xcolor}
\usepackage[compress,numbers,sort]{natbib}
\usepackage{colortbl}
\usepackage{pdflscape}
\usepackage{color}
\usepackage{mathtools}
\usepackage{colortbl}
\usepackage{comment}
\definecolor{Gray}{gray}{0.95}
\definecolor{RGray}{gray}{0.85}
\definecolor{CGray}{gray}{0.93}
\definecolor{piggypink}{rgb}{0.99, 0.87, 0.9}
\definecolor{babyblue}{rgb}{.67,.83,.99}

\newcommand{\B}{{\cal B}}
\newcommand{\W}{{\cal W}}

\newcommand{\SU}{{\rm SU}}

\newcommand{\U}{{\rm U}}

\newcommand{\V}{{\cal V}}

\renewcommand{\L}{{\cal L}}

\definecolor{nicered}{rgb}{0.7,0.1,0.1}
\definecolor{nicegreen}{rgb}{0.1,0.5,0.1}
\definecolor{red}{rgb}{1.0, 0, 0}
\definecolor{niceblue}{rgb}{0,0,0.8}
\definecolor{red}{rgb}{1.0, 0, 0}
\hypersetup{colorlinks,bookmarksopen,bookmarksnumbered,
linkcolor=blus,pdfstartview=FitH,urlcolor=rossos,citecolor=verde}
\allowdisplaybreaks

\definecolor{rosso}{cmyk}{0,1,1,0.4}
\definecolor{rossos}{cmyk}{0,1,1,0.55}
\definecolor{rossoc}{cmyk}{0,1,1,0.2}
\definecolor{blu}{cmyk}{1,1,0,0.3}
\definecolor{blus}{cmyk}{1,1,0,0.6}
\definecolor{bluc}{cmyk}{1,1,0,0.1}
\definecolor{verde}{cmyk}{0.92,0,0.59,0.25}
\definecolor{verdec}{cmyk}{0.92,0,0.59,0.15}
\definecolor{verdes}{cmyk}{0.92,0,0.59,0.4}

\def\eq#1{{Eq.~(\ref{#1})}}
\def\eqs#1#2{{Eqs.~(\ref{#1})--(\ref{#2})}}
\def\fig#1{{Fig.~\ref{#1}}}

\def\Table#1{{Table~\ref{#1}}}

\def\sect#1{{Sect.~\ref{#1}}}

\def\vev#1{\left\langle #1\right\rangle}

\renewcommand{\bar}{\overline}

\newcommand{\beq}{\begin{equation}}
\newcommand{\eeq}{\end{equation}}
\newcommand{\bea}{\begin{eqnarray}}
\newcommand{\eea}{\end{eqnarray}}

\renewcommand{\[}{\left[}
\renewcommand{\]}{\right]}
\renewcommand{\(}{\left(}
\renewcommand{\)}{\right)}

\renewcommand{\S}{\mathcal{S}}

\def\be{\begin{equation}}
\def\ee{\end{equation}}

\usepackage{xcolor}

\begin{document}

\begin{center}  
{\LARGE
\bf\color{blus} 
Higgs physics confronts the $M_W$ anomaly
} \\
\vspace{0.8cm}

{\bf Luca Di Luzio, Ramona Gr\"ober, Paride Paradisi }\\[7mm]

{\it Dipartimento di Fisica e Astronomia `G.~Galilei', Universit\`a di Padova, Italy}\\[1mm]
{\it Istituto Nazionale Fisica Nucleare, Sezione di Padova, Italy}\\[1mm]

\vspace{0.3cm}
\begin{quote}

The recent high-precision measurement of the $W$ mass by the CDF collaboration is in sharp tension with the Standard Model prediction as obtained by the electroweak fit. If confirmed, this 
finding can only be explained in terms of new physics effects.
In this work, we point out a 
generic
connection between
the $M_W$ anomaly 
and Higgs physics observables such as $h\to\gamma\gamma, Z\gamma$ and the ratio $h\to ZZ/WW$. Moreover, 
we systematically classify new physics 
scenarios which can address the $M_W$ anomaly via a tree-level 
contribution to the $\hat T$ parameter. 
These include a real scalar triplet, 
a scalar quadruplet with the same hypercharge of the Higgs doublet, 
a $Z'$ boson, a vector triplet with unit hypercharge  
and 
a vector boson with the gauge quantum numbers of the Higgs 
doublet. 
These solutions to the $M_W$ anomaly are characterized by 
new physics states 
which are typically too heavy to be discovered in direct searches, but which might leave 
their imprints 
in Higgs physics. 

\end{quote}
\thispagestyle{empty}
\end{center}

\bigskip
\tableofcontents

\clearpage

\section{Introduction}
\label{sec:intro}

Electroweak precision observables 
have played a crucial role to firmly establish the Standard Model (SM) as a quantum field theory 
and to constrain possible New Physics (NP) extensions.
Remarkably, electroweak precision observables were instrumental to indirectly infer 
the masses of the top quark and the Higgs boson well before their direct detection at Tevatron and LHC, respectively.
Nowadays, the SM electroweak fit is performed using as input parameters the fine structure constant $\alpha$, the muon decay constant $G_\mu$, the $Z$ boson mass $M_Z$, the strong coupling $\alpha_s(M_Z)$, the top quark mass $m_t$, the Higgs mass $M_h$, and the hadronic contribution to the running of $\alpha$, i.e.~$\Delta\alpha^{5}_{\rm had}(M_Z)$. In terms of these parameters, all other observables can be predicted.
In particular, the resulting value of the $W^\pm$ boson mass from the electroweak fit is $M_W = 80354.5 \pm 5.7\,\text{MeV}$ \cite{deBlas:2021wap}.

The CDF collaboration has recently published a high-precision measurement of $M_W = 80433.5 \pm 9.4 \, \text{MeV}$ \cite{CDF:2022hxs}, whose precision exceeds that of the current PDG world average, $M_W=80379 \pm 12 \,\text{MeV}$ \cite{ParticleDataGroup:2020ssz}, obtained from the combination of all previous measurements from LEP, D0, CDF, and ATLAS. The new CDF value turns out to be considerably larger than the current 
PDG world average as well as the value previously inferred from the SM electroweak fit  \cite{deBlas:2021wap}. 

Taking the new CDF result at face value, 
a few collaborations have already assessed its impact in the global electroweak fit, in the attempt of highlighting the favoured NP scenario to solve this anomaly 
(see e.g.~\cite{Lu:2022bgw,Athron:2022qpo,Strumia:2022qkt,deBlas:2022hdk}). In particular, it turned out that universal NP models,
which are fully described by the famous $\hat S, \hat T,U,W,\text{and}\,Y$
parameters~\citep{Peskin:1990zt,Peskin:1991sw,Barbieri:2004qk},
provide an overall good quality of the fit.
The viable solutions prefer either $\hat T \approx 10^{-3}$ and $\hat S=U=W=Y=0$ or highly-correlated positive $\hat S$ and $\hat T$ parameters 
of comparable size $\hat T \sim \hat S \sim 10^{-3}$ and $U=W=Y=0$. If $\hat S$ and $\hat T$ are loop-induced, they are of order $(g^4_{\rm NP}/16\pi^2)\times M^2_W/M^2_{\rm{NP}}$ and therefore, weakly-interacting theories require $M_{\rm NP}$ to lie at the electroweak scale to accommodate the $M_W$ anomaly. 
Such a solution can be hardly reconciled with the direct-search bounds on new particles. Instead, tree-level NP effects --which are equivalent 
to the effects stemming from a strongly-coupled sector with $g_{\rm{NP}}\sim 4\pi$-- can provide the desired values of 
$\hat S$ and $\hat T$ even for $M_{\rm{NP}} \sim 10$ TeV.

The primary goal of this work is to establish a 
connection between the NP effects entering $M_W$ and Higgs physics observables.
Indeed, since 
within the SM Effective Field Theory (SMEFT)
$\hat S$ and $\hat T$ 
receive contributions respectively from
the $d=6$ operators $(H^\dagger \tau^a H) W^a_{\mu\nu}B^{\mu\nu}$ and 
$(H^\dag D_\mu H)((D_\mu H)^\dag H)$,
it seems rather natural that NP effects in $M_W$ need 
to be accompanied by modifications of the SM predictions for Higgs decay processes like $h \to \gamma\gamma,Z\gamma$ and $h \to ZZ,WW$. 
In \sect{sec:EFT} 
we quantitatively assess this connection 
in the context of the SMEFT. 
Another goal of the present analysis is to systematically classify explicit NP extensions of the SM which can give a sizeable contribution to $\hat T$ at the tree level. 
We provide this classification in \sect{sec:hatTatTree} 
and, for those simplified models predicting a positive shift in $\hat T$, we  
discuss accordingly the correlated signals in Higgs physics. 
We conclude in \sect{sec:conclusions} with a summary of our findings.

\section{SMEFT approach to the $M_W$ anomaly and Higgs physics}
\label{sec:EFT}

Parametrizing the SMEFT Lagrangian as 
\beq 
\mathcal{L}_{\rm SMEFT} = \mathcal{L}_{\rm SM} 
+ \sum_i c_i \mathcal{O}_i \, ,
\eeq
where we adopt the Warsaw basis \cite{Grzadkowski:2010es} 
and focus in particular on the following subset of operators, 
which are relevant for 
electroweak and Higgs physics:
\begin{align}
\mathcal{O}_{HW} &= (H^\dag H) W^a_{\mu\nu} W^{a\mu\nu} \, , \\
\mathcal{O}_{HB} &= (H^\dag H) B_{\mu\nu} B^{\mu\nu} \, , \\
\mathcal{O}_{HWB} &= (H^\dag \tau^a H) W^a_{\mu\nu} B^{\mu\nu} \, , \\
\mathcal{O}_{HD} &= (H^\dag D_\mu H)((D_\mu H)^\dag H) \, , \\
\mathcal{O}_H &= (H^\dag H)^3 \, , \\
\mathcal{O}_{H\Box} &= (H^\dag H) \Box (H^\dag H) \, , \\
\mathcal{O}_{eH} &= (H^\dag H) \bar \ell_L e_R H \, , \\ 
\mathcal{O}_{uH} &= (H^\dag H) \bar q_L u_R \tilde H \, , \\
\mathcal{O}_{dH} &= (H^\dag H) \bar q_L d_R H \, , 
\end{align}
with the covariant derivative defined as
$D_{\mu}=\partial_{\mu}+i g_2 W^{a}_{\mu}\tau^{a}+i g_1 B_{\mu} Y$. 
Employing the notation of Refs.~\cite{Barbieri:2004qk,Strumia:2022qkt}, 
the leading electroweak oblique corrections 
are described by\footnote{In this notation 
the
$S$ and $T$ parameters of Refs.~\cite{Peskin:1990zt,Peskin:1991sw} 
read
$S = 4 s^2_W \hat S / \alpha$ and $T = \hat T / \alpha$.} 
\begin{align} 
\label{eq:Stoc}
\hat S &\equiv \frac{c_W}{s_W} \Pi'(0)_{W_3B} = \frac{c_W}{s_W} v^2 c_{HWB} \, , \\
\label{eq:Ttoc}
\hat T &\equiv \frac{1}{M^2_W} (\Pi_{W_3W_3}(0) - \Pi_{W^+W^-}(0)) = - \frac{v^2}{2} c_{HD} \, , 
\end{align}
with $v = 246$ GeV and   
$s_W \equiv \sin\theta_W$ 
($c_W \equiv \cos\theta_W$). 
We remark that in \eqs{eq:Stoc}{eq:Ttoc} we only included 
so-called ``universal'' bosonic operators. 
Upon applying the equations of motion in a given basis, 
other fermionic operators can contribute as well to the $\hat S$ and $\hat T$ parameters (see e.g.~\cite{Elias-Miro:2013eta, Wells:2015uba}). 
Concretely, in terms of the Warsaw basis these are four-fermion operators as well as operators of the type $(H \overleftrightarrow{D}_{\mu} H)( \bar{\psi} \gamma^{\mu}\psi)$. These operators can also lead to contributions to electroweak precision observables beyond the oblique parameters 
(with the exception of top-quark operators\footnote{Top-quark operators can be (weakly) constrained 
by top-quark physics \cite{Hartland:2019bjb} and 
via their loop contributions 
by electroweak 
observables \cite{Dawson:2022bxd} and Higgs physics \cite{Alasfar:2022zyr}.}) 
and hence are neglected in the present analysis.

The $M_W$ anomaly could be due to a universal new physics correction 
to $\hat T$ \cite{Strumia:2022qkt}
\beq 
\hat T \simeq (0.84 \pm 0.14) \times 10^{-3} \, , 
\eeq
($c_{HD} = - (0.17 \pm 0.07 / \text{TeV})^2$) 
as well as a correlated contribution to 
$\hat S \sim 10^{-3}$ 
($c_{HWB} \sim (0.07 / \text{TeV})^2$)
of the same size of $\hat T$, but 
compatible with zero \cite{Strumia:2022qkt,deBlas:2022hdk}. 
The inclusion of higher-order corrections 
in the momentum expansion of the inverse propagators 
($Y$ and $W$) does not alter significantly the fit \cite{Strumia:2022qkt}, while  
a non-vanishing $\hat U$ parameter 
can also explain by itself
the $M_W$ anomaly \cite{deBlas:2022hdk}.   
However, under the assumption of heavy NP,  
which is captured by the SMEFT 
description, 
the $\hat U$ 
parameter is usually neglected 
since it arises from $d=8$ operators. 

\begin{figure}[t]
\centering
\vspace{-0.5cm}
\includegraphics[width=15cm]{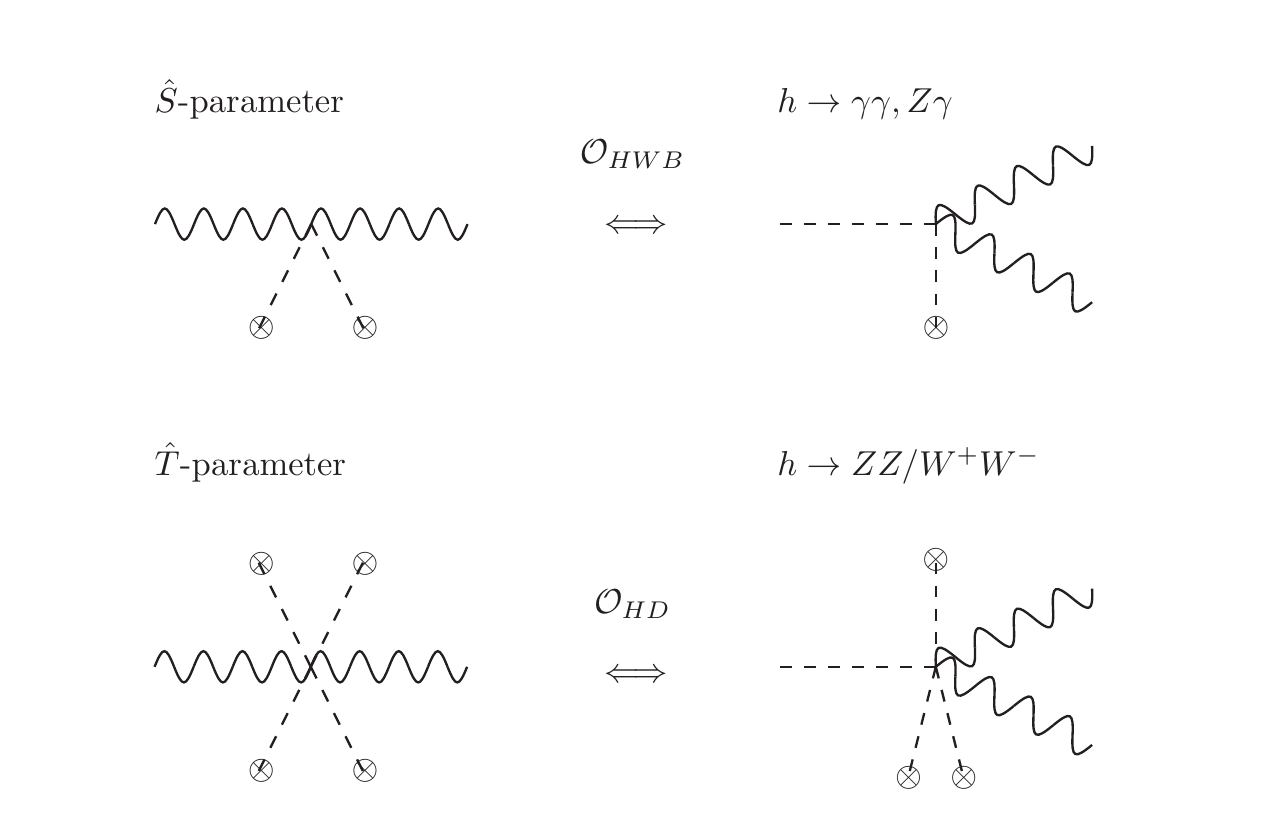}
\caption{$\hat S$ and $\hat T$ vs.~Higgs connection.}
\vspace{0.5cm}
\label{fig:STtoHiggs}
\end{figure}

Since the $\hat S$ and $\hat T$ parameters are obtained 
by condensing the Higgs fields in $\mathcal{O}_{HWB}$ 
and $\mathcal{O}_{HD}$, there is 
clearly a  
connection with Higgs physics, 
as highlighted schematically in \fig{fig:STtoHiggs}. 

Writing the SMEFT Lagrangian in the 
electroweak broken phase as 
\begin{align} 
\mathcal{L}^{\rm int}_{\rm SMEFT} 
\label{eq:SMEFTbroken}
&\ni g_{hWW}^{(1)} h W^+_\mu W^{-\mu} + g_{hWW}^{(2)}h W^+_{\mu\nu}W^{-\mu\nu}
+ g_{hZZ}^{(1)} h Z_\mu Z^{\mu}+g_{hZZ}^{(2)} h Z_{\mu\nu}Z^{\mu\nu} +g_{h\gamma\gamma } h F_{\mu\nu} F^{\mu\nu}  \nonumber \\ & +g_{h\gamma Z } h F_{\mu\nu} Z^{\mu\nu} 
+ g_{hhh} h^3 
+ (g_{he} h \bar e_L e_R 
+ g_{hu} h \bar u_L u_R
+ g_{hd} h \bar d_L d_R + \text{h.c.}) + \ldots \, ,
\end{align}
one finds at tree level (see e.g.~\cite{Dedes:2017zog})
\begin{align}
    g_{hWW}^{(1)} &= \frac{2M_W^2}{v} \left( 1 - \frac{v^2}{4} (c_{HD} - 4 c_{H\Box}) \right) \, ,
    \\
     g_{hWW}^{(2)}&=
     2 v c_{HW} \, , \\
   g_{hZZ}^{(1)} &= \frac{M^2_Z}{v} \left( 1 + \frac{v^2}{4} (c_{HD} + 4 c_{H\Box}) \right) \, , \\
   g_{hZZ}^{(2)}&=
    v  \left[ \frac{M_W^2}{M_Z^2}c_{HW}+ \frac{M_Z^2-M_W^2}{M_Z^2}c_{HB}+\frac{g_1 g_2  }{g_1^2+g_2^2} c_{HWB} \right] \, , \\
   g_{h\gamma\gamma}&=v \left[ \frac{M_Z^2-M_W^2}{M_Z^2} c_{HW}+\frac{M_W^2}{M_Z^2} c_{HB}- \frac{g_1 g_2}{g_1^2+g_2^2}c_{HWB} \right]  \, , \\
    g_{h\gamma Z}&= 2 v \left[ \frac{g_1 g_2}{g_1^2+g_2^2} c_{HW}- \frac{g_1 g_2}{g_1^2+g_2^2} c_{HB}+\frac{1}{2} \frac{g_1^2- g_2^2}{g_1^2+g_2^2}c_{HWB} \right] \, , \\
   \label{eq:ghhh}
    g_{hhh} &= - \frac{M^2_h}{2v} 
    \left[ 1 - \frac{3 v^2}{4} (c_{HD} - 4 c_{H\Box}) -\frac{2 v^4}{M_h^2} c_H\right]  \, , \\
    g_{h\psi} &= -\frac{m_\psi}{v} \left(1 - \frac{v^2}{4}(c_{HD} - 4 c_{H\Box})\right) + \frac{c_{\psi H}v^2}{\sqrt{2}} \, ,
\end{align}
with $\psi = e,u,d$. 
In order to canonically normalize the Higgs kinetic term we adopted the field redefinition
\begin{equation}
    h\to h \left(1+ v^2 (c_{H\Box}-\frac{1}{4}c_{HD}) \left(1+\frac{h}{v}+\frac{h^2}{3 v^2}\right) \right)  \, ,
\end{equation}
which removes the momentum-dependence in the Higgs self-couplings.

The most relevant Higgs observables, which are affected by the presence of non-zero $\hat S, \hat T \sim 10^{-3}$, 
are the Higgs boson decays into vector bosons. 
Defining the Higgs signal strengths as\footnote{For illustration of the argument, we work under the assumption that the Higgs coupling to gluons 
is not modified by NP and neglect other production channels than the ones to gluons.} 
\beq 
\mu_{VV'} \equiv \frac{\Gamma(h \to VV')}{\Gamma^{\rm SM}(h \to VV')} \, , 
\eeq
with $V,V' = \gamma,Z,W$, 
we can compute the corrections arising 
from the modified Higgs couplings in 
\eq{eq:SMEFTbroken}. 
For the observables related to $\hat S$ ($c_{HWB}$)
one finds 
(see e.g.~\cite{Grojean:2013kd})
\begin{align} 
\mu_{\gamma\gamma}  
&\simeq 
1 + \frac{4\pi v^2}{I^\gamma \alpha} 
(s^2_W c_{HW} + c^2_W c_{HB} - s_W c_W c_{HWB}) \nonumber \\
&\simeq 1 + 0.23  \( \frac{\hat S}{10^{-3}} \) \, , \\
\mu_{Z\gamma} 
&\simeq 
1 + \frac{4\pi v^2}{I^Z \alpha} 
(s_W c_W c_{HW} - s_W c_W c_{HB} - \frac{1}{2} (c^2_W - s^2_W) c_{HWB})  \nonumber \\
&\simeq 1 + 0.084  \( \frac{\hat S}{10^{-3}} \) \, , 
\end{align}
where in the last steps we used the SM values $I^\gamma = -1.64$ and $I^Z = -2.84$, 
and neglected the contribution of $c_{HW}$ and $c_{HB}$ 
in order to highlight the connection with $\hat S$. 
Hence, a $+20\%$ modification 
of $\mu_{\gamma\gamma}$ is generically expected 
for $\hat S \sim 10^{-3}$, 
which is in the ballpark of the present 
LHC experimental 
sensitivity at the $10\%$ level \cite{ATLAS:2019nkf}. 
On the other hand, the predicted shift in $\mu_{Z\gamma}$ 
is presently too small to be detected, although an $\mathcal{O}(10\%)$ 
sensitivity might be achieved at the HL-LHC \cite{Cepeda:2019klc}. 
Another observable that is directly 
sensitive to $\hat T$ via $c_{HD}$ is the ratio 
$\mu_{ZZ} / \mu_{WW}$ \cite{Farina:2012ea}, 
which reads 
\beq 
\frac{\mu_{ZZ}}{\mu_{WW}} \simeq 1 + 2 c_{HD} v^2 \simeq 
1 - 0.0034 \( \frac{\hat T}{0.84 \times 10^{-3}} \) \, .  
\eeq 
However, also in this case the predicted deviation is too small to be presently detected.

\section{Heavy new physics: tree-level contributions to $\hat T$}
\label{sec:hatTatTree}

Since the $M_W$ anomaly 
hints at 
a sizeable $\hat T \sim 10^{-3}$, correlated with an $\hat S$ 
parameter compatible with zero \cite{Strumia:2022qkt,deBlas:2022hdk},  
we will now focus on 
heavy NP extensions which can yield 
a tree-level 
contribution to $\hat T$ 
via the operator $\mathcal{O}_{HD}$.\footnote{We note that in a few models discussed below ($\Delta_1$, $\B$, $\W$ and $\L$ from Table~\ref{tab:hatTatTreeLevel}) 
one also generates fermionic operators that affect $\hat{S}$ and $\hat{T}$ (cf.~discussion below \eq{eq:Ttoc}). 
However, these contributions are always proportional to the couplings of the new particle
to SM fermions and therefore they can be parametrically suppressed.} 

It turns out that such states are either scalars $\S$ or vectors $\V$, whose 
quadratic
Lagrangian 
can be written as 
\begin{align}
\mathcal{L}^{\rm quad}_\S &= \eta [(D_\mu \S)^\dag D^\mu \S - M^2_\S \S^\dag \S] \, , \\
\mathcal{L}^{\rm quad}_\V &= \eta [(D_\mu \V_\nu)^\dag D^\nu \V^\mu 
- (D_\mu \V_\nu)^\dag D^\mu \V^\nu + M^2_\V \V^\dag_\mu \V^\mu] \, ,
\end{align} 
with the prefactor $\eta = 1$ ($\eta = 1/2$) for a complex (real) representation. 
The representations which can generate $\mathcal{O}_{HD}$, or higher-dimensional variants thereof such as $(H^\dag H) \mathcal{O}_{HD}$,
at tree-level are displayed in \Table{tab:hatTatTreeLevel}
(for similar classifications see also Refs.~\cite{Henning:2014wua,Dawson:2017vgm,Corbett:2017ieo,deBlas:2017xtg}). 
In the following, we discuss in detail each simplified model, 
and for those cases leading to a positive $\hat T$ 
we analyze in turn the correlated signals in Higgs physics. 
\begin{table}[!ht]
	\centering
	\begin{tabular}{|c|c|c|c|c|c|c|}
	\rowcolor{CGray} 
	\hline
	Field & Spin & $\SU(3)_C$ & $\SU(2)_L$ & $\U(1)_Y$ & $\text{sign}(\hat T)$ & $\hat S$ \\ 
	\hline 
    \rowcolor{piggypink} 
	$\Delta$ & $0$ & 1 & 3 & 0 & $+$ & $\times$ \\ 
	$\Delta_1$ & $0$ & 1 & 3 & 1 & $-$ & $\times$ \\
	\rowcolor{piggypink} 
	$\Theta_1$ & $0$ & 1 & 4 & 1/2 & $+$ & $\times$ \\ 
	$\Theta_3$ & $0$ & 1 & 4 & 3/2 & $-$ & $\times$ \\ 
	\rowcolor{piggypink} 
	$\B$ & $1$ & 1 & 1 & 0 & $+$ & $\times$ \\ 
	$\B_1$ & $1$ & 1 & 1 & 1 & $-$ & $\times$ \\ 
	$\W$ & $1$ & 1 & 3 & 0 & $-$ & $\times$ \\ 
	\rowcolor{piggypink}
	$\W_1$ & $1$ & 1 & 3 & 1 & $+$ & $\times$ \\ 
	$\L$ & $1$ & 1 & 2 & 1/2 & $+/-$ & $\checkmark$ \\ 
	\hline
	\end{tabular}	
	\caption{\label{tab:hatTatTreeLevel} 
	New physics states which can yield a tree-level contribution to $\hat T$ via $d \leq 4$ interactions 
	with SM states. Highlighted in pink are the representations predicting a positive shift on $\hat T$. The last column indicates whether a 
	tree-level contribution to $\hat S$ 
	is generated $(\checkmark)$ or not $(\times)$.}
\end{table}

\subsection{$\Delta \sim (1,3,0)_\S$}
From the interaction Lagrangian 
\beq 
\mathcal{L}_{\Delta}^{\rm int} \ni - \kappa_\Delta H^\dag \Delta^a \sigma^a H - \frac{\lambda_{H\Delta}}{2} 
(H^\dag H) \Delta^a \Delta^a
\, , 
\eeq
one obtains
\beq 
c_{HD} = - 2 \frac{\kappa^2_\Delta}{M_\Delta^4} \, , 
\eeq
and hence
\beq 
\label{eq:hatTDelta}
\hat T = \frac{\kappa^2_\Delta v^2}{M_\Delta^4} 
= 0.84 \times 10^{-3} 
\(\frac{|\kappa_{\Delta}|}{M_{\Delta}}\)^2 \( \frac{8.5 \, \text{TeV}}{M_{\Delta}} \)^2 \, ,
\eeq
which has the correct sign to explain the 
$M_W$ anomaly.

When the $\Delta$ is not integrated out, 
the tree-level contribution to $\hat T$ 
can be alternatively understood to arise from 
the generation of a 
tree-level vacuum expectation value (VEV) for $\Delta$, 
that is $\vev{\Delta} \equiv v_\Delta = \kappa_\Delta v^2 / (2 M^2_\Delta)$.
In general, the VEV of a scalar representation $\S \sim (1,2j+1,y)$ yields \cite{Gunion:1989we,DiLuzio:2015oha}
\beq 
\label{eq:VEVhatT}
    \hat T \simeq  
    4\(\eta [j (j+1) - y^2] -2 y^2 \) \frac{\vev{\S}^2}{v^2}  
    \, ,
    \eeq
with $\vev{\S} = \alpha v_\S$, where $\alpha = 1$ ($\alpha = 1/\sqrt{2}$) 
for a real (complex) representation and 
$v_\S$ 
is the VEV of the canonically normalized real scalar component of $\S$.
In the case $\S = \Delta$ this yields 
$\hat T \simeq \kappa^2_\Delta v^2 / M_\Delta^4$, as in \eq{eq:hatTDelta}.
 
The connection between this scalar triplet and electroweak 
precision measurements was previously considered e.g.~in Refs.~\cite{Lynn:1990zk,Chen:2006pb,Bandyopadhyay:2020otm}. 
Note that the perturbativity range of the massive $\kappa_{\Delta}$ parameter can be obtained 
by requiring that finite loop corrections to the trilinear scalar vertex $\Delta H^\dag H$ remain smaller than 
the tree-level value \cite{DiLuzio:2016sur}. 
This yields $|\kappa_{\Delta}| / M_{\Delta} \lesssim 4 \pi$ \cite{DiLuzio:2017tfn}. 
Hence, a scalar triplet well above the TeV scale and with perturbative couplings
can explain the value of $\hat T$ while 
easily evading all direct collider searches. 
In particular, saturating the perturbativity bound, it turns out that $M_{\Delta} \lesssim 100$ TeV.

Other coefficients which are unavoidably generated after integrating out $\Delta$ are directly correlated with 
$\hat T$ via the coupling 
$\kappa_{\Delta}$: 
\begin{align}
c_H &= - 4 \frac{\kappa_{\Delta}^2}{M_{\Delta}^4} 
\(\frac{\lambda_{H\Delta}}{8} - \lambda\) 
= - 4 \frac{\hat T}{v^2} 
\(\frac{\lambda_{H\Delta}}{8} - \lambda\) 
\, , \\
c_{H\Box} &= \frac{\kappa_{\Delta}^2}{2 M_{\Delta}^4} 
= \frac{\hat T}{2 v^2}
\, , \\  
c_{eH,\, uH,\, dH} &= \frac{\kappa_{\Delta}^2 Y_{e,u,d}}{M_{\Delta}^4} 
= \frac{\hat T}{v^2} Y_{e,u,d} \, ,
\end{align}
where $\lambda$ is the SM quartic Higgs coupling
and $Y_{e,u,d}$ are SM Yukawas. 
Hence, in the triplet model a non-zero $\hat T$ 
can be correlated to various Higgs signals 
(see \eq{eq:SMEFTbroken}).
The strongest dependence on $\hat{T}$ is via the modification of the trilinear Higgs self-coupling, but one needs to keep in mind that the prospect for 
its
measurement at the HL-LHC is $0.1<g_{hhh}/g_{hhh}^{\rm SM}<2.3$ \cite{DiMicco:2019ngk} while the model predicts (setting $\lambda_{H\Delta}=0$) deviations of $\mathcal{O}(1\%)$ in $g_{hhh}$ (cf.~\eq{eq:ghhh}). Similarly, the deviations in the $W$ and $Z$ couplings are out of reach for the LHC.

\subsection{$\Delta_1 \sim (1,3,1)_\S$}
From the interaction Lagrangian
\beq 
\label{eq:LintDelta1}
\mathcal{L}^{\rm int}_{\Delta_1} \ni - \kappa_{\Delta_1} (\Delta_1^a)^\dag \tilde H^\dag  \sigma^a H + \text{h.c.}
\, , 
\eeq
one obtains
\beq 
c_{HD} = 4 \frac{|\kappa_{\Delta_1}|^2}{M_{\Delta_1}^4} \, ,
\eeq
and hence
\beq 
\hat T = - 2 \frac{|\kappa_{\Delta_1}|^2 v^2}{M_{\Delta_1}^4} \, , 
\eeq
which predicts the wrong sign to explain the new CDF $M_W$ value.

\subsection{$\Theta_1 \sim (1,4,1/2)_\S$}

Electroweak quadruplets contribute to $\hat T$ via the $d=8$ operator 
$(H^\dag H) \mathcal{O}_{HD}$.\footnote{Note that quadruplets do not generate 
tree-level contributions to the $\hat S$ and $\hat U$ parameters at $d=8$ \cite{Murphy:2020rsh}.}
It is hence more practical to 
directly
compute 
$\hat T$ via the VEV contribution. 
We consider the interaction Lagrangian
\beq 
\mathcal{L}^{\rm int}_{\Theta_1} \ni 
M^2_{\Theta_1} (\Theta_1)_{ijk} (\Theta^*_1)^{ijk} 
- \lambda_{H3\Theta_1} H^{*i} (\Theta_1)_{ijk} 
H^{*j} \epsilon^{kl} H_l + \text{h.c.}
\, , 
\eeq
in a phase convention where $\lambda_{H3\Theta_1}$ is real and 
we employed a symmetric tensor notation for the quadruplet, with  
latin indices in $\SU(2)_L$ space and 
$\epsilon = i \sigma^2$. 
The embedding of the 
canonically normalized
charge 
eigenstates reads
$(\Theta_1)_{111} = \Theta^{++}_1$, 
$(\Theta_1)_{112} = \frac{1}{\sqrt{3}} \Theta^{+}_1$, 
$(\Theta_1)_{122} = \frac{1}{\sqrt{3}} \Theta^{0}_1$, 
$(\Theta_1)_{222} = \Theta^{-}_1$. 
In particular, for the VEV of the neutral component one obtains
\beq 
\vev{\Theta^0_1} \equiv \frac{v_{\Theta_1}}{\sqrt{2}} \simeq \frac{\lambda_{H3\Theta_1}  v^3}{2\sqrt{6} M^2_{\Theta_1}}  \, , 
\eeq
and hence, using the VEV formula in \eq{eq:VEVhatT}
\beq 
\hat T \simeq 12 \frac{\vev{\Theta^0_1}^2}{v^2} 
\simeq 
\frac{\lambda_{H3\Theta_1}^2 v^4}{2M^4_{\Theta_1}} 
= 0.84 \times 10^{-3} \, \lambda_{H3\Theta_1}^2 \(\frac{1.2 \, \text{TeV}}{M_{\Theta_1}}\)^4
\, , 
\eeq
which has the correct sign to explain the $M_W$ anomaly.  
Note that due to the different scaling of the $\hat T$ parameter (compared e.g.~to the triplet case in 
\eq{eq:hatTDelta}) the mass of the quadruplet needs to be around 1 TeV for 
$\mathcal{O}(1)$
couplings to the Higgs. 

At the leading order in the SMEFT, one also generates the 
Wilson coefficient  
\beq
c_H = \frac{\lambda_{H3\Theta_1}^2}{M^2_{\Theta_1}}
= \frac{2 M^2_{\Theta_1}}{v^4} \hat T 
= \frac{0.040}{v^2} \( \frac{M_{\Theta_1}}{1.2 \,\text{TeV}} \)^2 
\( \frac{\hat T}{0.84 \times 10^{-3}} \)
\, ,
\eeq
which is directly correlated with 
$\hat T$. 
Substituting the above value of $c_H$ into \eq{eq:ghhh} we then obtain 
\beq 
\frac{g_{hhh}}{g^{\rm SM}_{hhh}} - 1 =  - 31\% \( \frac{M_{\Theta_1}}{1.2 \,\text{TeV}} \)^2 
\( \frac{\hat T}{0.84 \times 10^{-3}} \) \, ,
\label{eq:trilinear_Higgs}
\eeq 
which implies up to an $\mathcal{O}(1)$ variation in the trilinear Higgs self-coupling, 
depending on the value of $M_{\Theta_1}$, which receives an upper bound from perturbativity 
(see e.g.~\cite{Dawson:2017vgm}). 
All in all, the reason for such a large effect in the trilinear Higgs self-coupling 
 can be understood from the fact that in the quadruplet case $\hat T$ is 
generated by a $d=8$ operator, while the contribution to $c_H$ arises at $d=6$. 

\subsection{$\Theta_3 \sim (1,4,3/2)_\S$}

We consider the interaction Lagrangian
\beq 
\mathcal{L}^{\rm int}_{\Theta_3} \ni 
M^2_{\Theta_3} (\Theta_3)_{ijk} (\Theta^*_3)^{ijk}
- \lambda_{H3\Theta_1} H^{*i} H^{*j} H^{*k} (\Theta_3)_{ijk}  + \text{h.c.}
\, , 
\eeq 
in a phase convention where $\lambda_{H3\Theta_3}$ is real. 
The 
embedding of the 
canonically normalized
charge 
eigenstates reads
$(\Theta_3)_{111} = \Theta^{+++}_3$, 
$(\Theta_3)_{112} = \frac{1}{\sqrt{3}} \Theta^{++}_3$, 
$(\Theta_3)_{122} = \frac{1}{\sqrt{3}} \Theta^{+}_3$, 
$(\Theta_3)_{222} = \Theta^{0}_3$. 
For the VEV of the neutral component one obtains
\beq 
\vev{\Theta^0_3} \equiv \frac{v_{\Theta_3}}{\sqrt{2}} \simeq \frac{\lambda_{H3\Theta_3}  v^3}{2\sqrt{2} M^2_{\Theta_3}}  \, , 
\eeq
and hence, using the VEV formula in \eq{eq:VEVhatT}
\beq 
\hat T \simeq - 12 \frac{\vev{\Theta^0_3}^2}{v^2} 
\simeq 
- \frac{3 \lambda_{H3\Theta_3}^2 v^4}{2M^4_{\Theta_3}} \, ,
\eeq
which predicts the wrong sign to solve the $M_W$ anomaly.

\subsection{$\B \sim (1,1,0)_\V$}
From the interaction Lagrangian 
\beq 
\mathcal{L}^{\rm int}_{\B} \ni - g_\B^H \B^\mu H^\dag i D_\mu H + \text{h.c.} \, , 
\eeq
one obtains 
\beq 
c_{HD} = - 2 \frac{(\text{Re}\, (g_\B^H))^2}{M^2_{\B}} \, , 
\eeq
and hence
\beq 
\label{eq:thatB}
\hat T = \frac{(\text{Re}\, (g_\B^H))^2 v^2}{M^2_{\B}} 
= 0.84 \times 10^{-3} \,
(\text{Re}\, (g_\B^H))^2 \( \frac{8.5 \, \text{TeV}}{M_{\B}} \)^2 \, ,
\eeq
which has the correct sign to explain the 
$M_W$ anomaly. 
The possibility of raising the $M_W$ mass via a $Z'$ 
boson was previously considered e.g.~in 
\cite{Alguero:2022est,Strumia:2022qkt}. 
Note that the $Z'$ phenomenology
highly depends on its 
coupling to SM fermions, which have not been specified here.

\subsection{$\B_1 \sim (1,1,1)_\V$} 
From the interaction Lagrangian
\beq 
\label{eq:LintB1}
\mathcal{L}^{\rm int}_{\B_1} \ni - g_{\B_1}^H \B_1^{\mu\dag} i D_\mu H^T i \sigma^2 H + \text{h.c.} \, , 
\eeq
one obtains 
\beq 
c_{HD} =  \frac{|g_{\B_1}^H|^2}{M^2_{\B_1}} \, , 
\eeq
and hence 
\beq 
\hat T = - \frac{|g_{\B_1}^H|^2 v^2}{2 M^2_{\B_1}} \, , 
\eeq
which predicts the wrong sign to accommodate the $M_W$ anomaly. 

\subsection{$\W \sim (1,3,0)_\V$} 
From the interaction Lagrangian
\beq 
\mathcal{L}^{\rm int}_{\W} \ni - \frac{1}{2} g_\W^H \W^{\mu a} H^\dag \sigma^a i D_\mu H + \text{h.c.} \, , 
\eeq
one obtains
\beq 
c_{HD} = \frac{(\text{Im}\, (g_\W^H))^2}{2 M^2_{\W}} \, , 
\eeq
and hence 
\beq 
\hat T = -\frac{(\text{Im}\, (g_\W^H))^2 v^2}{4 M^2_{\W}} \, , 
\eeq
which predicts the wrong sign to accommodate the $M_W$ anomaly. 

\subsection{$\W_1 \sim (1,3,1)_\V$} 
From the interaction Lagrangian
\beq 
\label{eq:LintW1}
\mathcal{L}^{\rm int}_{\W_1} \ni - \frac{1}{2} g_{\W_1}^H \W_1^{\mu a\dag} i D_\mu H^T i \sigma^2 \sigma^a H + \text{h.c.} \, , 
\eeq
one obtains 
\beq 
c_{HD} =  - \frac{|g_{\W_1}^H|^2}{4M^2_{\W_1}} \, , 
\eeq
and hence 
\beq 
\hat T = \frac{|g_{\W_1}^H|^2 v^2}{8M^2_{\W_1}} = 0.84 \times 10^{-3} \,
|g_{\W_1}^H|^2 \( \frac{3.0 \, \text{TeV}}{M_{\W_1}} \)^2 \, ,
\eeq
which has the correct sign to explain the $M_W$ 
anomaly. 
Note that 
the interaction in \eq{eq:LintW1} 
does not necessarily arise 
from a renormalizable theory 
and here we have included only $d=4$ 
interactions with the SM fields. 

\subsection{$\L \sim (1,2,1/2)_\V$} 
From the interaction Lagrangian
\begin{align}
\mathcal{L}^{\rm int}_{\L} &\ni - (\gamma_\L \L^\dag D^\mu H + \text{h.c.}) 
- i g^B_\L \L^\dag_\mu \L_\nu B^{\mu\nu} 
- i g^W_\L \L^\dag_\mu \sigma^a \L_\nu W^{a\mu\nu}
- h^{(2)}_\L \L^\dag H H^\dag \L \nonumber \\
&- ( h^{(3)}_\L (\L^\dag H)^2 + \text{h.c.}) 
\, , 
\end{align}
after an Higgs field redefinition in order to 
have canonical kinetic terms (see \cite{deBlas:2017xtg}),
one obtains 
\beq 
c_{HD} = \frac{g_1 g^B_\L |\gamma_\L|^2}{M_\L^4} 
- \frac{h^{(2)}_\L |\gamma_\L|^2}{M_\L^4} 
+\frac{2 \text{Re}\, (h^{(3)}_\L \gamma^{*2}_\L )}{M_\L^4}
\, ,
\eeq
and hence 
\beq 
\hat T = - \frac{g_1 g^B_\L |\gamma_\L|^2 v^2}{2 M_\L^4} 
+ \frac{h^{(2)}_\L |\gamma_\L|^2 v^2}{2 M_\L^4} 
- \frac{\text{Re}\, (h^{(3)}_\L \gamma^{*2}_\L) v^2}{M_\L^4}
\, , 
\eeq
which can have both signs. 
Moreover, after integrating out $\L$, 
also the Wilson 
coefficient 
\beq 
c_{WB} = -\frac{g_1 g_2 |\gamma_\L|^2}{4 M_\L^4} 
-\frac{g_2 g^B_{\L}|\gamma_\L|^2}{4 M_\L^4} 
-\frac{g_1 g^W_{\L}|\gamma_\L|^2}{4 M_\L^4}
\eeq
is generated, 
which implies 
\beq 
\hat S = \frac{c_W}{s_W} \[  -\frac{g_1 g_2 |\gamma_\L|^2 v^2}{4 M_\L^4} 
-\frac{g_2 g^B_{\L}|\gamma_\L|^2 v^2}{4 M_\L^4} 
-\frac{g_1 g^W_{\L}|\gamma_\L|^2 v^2}{4 M_\L^4} \] \, ,
\eeq
which is of the same order as the contribution to the $\hat{T}$ parameter. Hence, this model could yield a large effect 
in the $h\to \gamma \gamma$ rate (as discussed in Sect.~\ref{sec:EFT}).

\section{Conclusions}
\label{sec:conclusions}

In this work, we have investigated the connection 
between the $M_W$ anomaly 
(stemming from the recent $M_W$ measurement by
the CDF collaboration~\citep{CDF:2022hxs}) 
and Higgs physics. 
This connection is quite natural from the 
point of view of the SMEFT, 
since the 
$\hat S$ and $\hat T$ parameters arise from 
$d=6$ operators containing multiple insertions of 
the Higgs doublet set on its VEV. 
Hence, by promoting one of those VEVs 
to a dynamical field, one automatically predicts 
modified Higgs signals (cf.~\fig{fig:STtoHiggs}). 
The largest effect turns out to be 
in $h \to \gamma\gamma$, which is modified up to 
$+20\%$ for values of $\hat S \sim 10^{-3}$ 
which are generically suggested by the electroweak fit. 
Modifications of the $h \to Z\gamma$ rate as 
well as the ratio $h \to ZZ /WW$ (which might be used to 
probe the $\hat T$ parameter) are instead too small 
to be presently detected. 

Since a non-zero and positive $\hat T \sim 10^{-3}$ 
is suggested by global electroweak fits, 
in \sect{sec:hatTatTree} 
we classified SM extensions which predict a 
positive tree-level shift of $\hat T$. 
Remarkably, there are only few solutions 
with NP states that couple directly to the Higgs 
via $d \leq 4$ interactions: 
a scalar triplet $\Delta \sim (1,3,0)_\S$, 
a scalar quadruplet $\Theta_1 \sim (1,4,1/2)_\S$, 
a $Z'$ boson $\B \sim (1,1,0)_\V$, 
a vector triplet $\W_1 \sim (1,3,1)_\V$ 
and a vector boson $\L \sim (1,2,1/2)_\V$. 
In all these cases the value of $\hat T \sim 10^{-3}$ 
can be easily explained via a NP state 
with mass around 10 TeV  
and $\mathcal{O}(1)$ couplings to the Higgs 
(barring the quadruplet case which needs to be at the TeV scale 
and implies as well a large deviation in the trilinear Higgs self-coupling -- see Eq.~(\ref{eq:trilinear_Higgs})). 

Although the NP states implied by such scenarios can escape direct detection at particle colliders, 
they might still leave their imprints via modifications of Higgs signals, 
which become 
especially
correlated in explicit models.

\section*{Acknowledgments} 

We thank 
David Marzocca,  
Alessandro Strumia and Nicol\`o Toniolo 
for useful discussions.


\bibliographystyle{utphys}
\bibliography{bibliography}


\end{document}